\documentclass[conference,10pt]{IEEEtran}
\usepackage{cite}
\usepackage{amsmath,amssymb,amsfonts}
\usepackage{algorithmic}
\usepackage{algorithm}
\usepackage{graphicx}
\usepackage{textcomp}
\usepackage{xcolor}
\usepackage{balance}
\usepackage{url}
 \usepackage{hyperref}
\usepackage{silence}
\usepackage{caption}
\begin{document}

\def\b{{\mathbf b}}
\def\c{{\mathbf c}}
\def\n{{\mathbf n}}
\def\v{{\mathbf v}}
\def\x{{\mathbf x}}
\def\y{{\mathbf y}}
\def\X{{\mathbf X}}
\def\Y{{\mathbf Y}}
\def\F{{\mathbf F}}
\def\G{{\mathbf G}}
\def\H{{\mathbf H}}
\def\W{{\mathbf W}}
\def\A{{\mathbf A}}
\def\B{{\mathbf B}}
\def\C{{\mathbf C}}
\def\M{{\mathbf M}}
\def\Z{{\mathbf Z}}
\def\U{{\mathbf U}}
\def\V{{\mathbf V}}
\def\I{{\mathbf I}}
\def\w{{\mathbf w}}
\def\S{{\mathbf \Sigma}}
\def\SigmaW{{\mathbf \sigma_\v}}
\def\L{{\cal L}}

\newcommand{\temp}[1]{\textcolor{red}{#1}}
\newcommand{\mario}[1]{\textcolor{green}{#1}}

\IEEEoverridecommandlockouts

\title{Federated Latent Space Alignment\\ for Multi-user Semantic Communications\vspace{-.7cm}}
\author{Giuseppe Di Poce$^{3,*}$, Mario Edoardo Pandolfo$^{1,2,*}$, Emilio Calvanese Strinati$^3$, and Paolo Di Lorenzo$^{2,4}$ \medskip \\
$^1$ DIAG Department, Sapienza University of Rome, via Ariosto 25, Rome, Italy.\\
$^2$ Consorzio Nazionale Interuniversitario per le Telecomunicazioni (CNIT), Parma, Italy.\\
$^3$ CEA Leti, University Grenoble Alpes, 38000, Grenoble, France.\\
$^4$ DIET Department, Sapienza University of Rome, Via Eudossiana 18, Rome, Italy. \smallskip\\
e-mail: \{marioedoardo.pandolfo, paolo.dilorenzo\}@uniroma1.it, \{giuseppe.dipoce, emilio.calvanese-strinati\}@cea.fr.
\thanks{This work was funded by the 6G-GOALS project (6G SNS-JU Horizon, n.101139232), the EU under NextGenerationEU RESTART (PE00000001), the SNS JU 6GARROW project (Horizon, n. 101192194), and the French government (France 2030 ANR, ref. 22-PEFT-0010). $^*$Equal contribution.}\vspace{-.3cm}}




\maketitle

\begin{abstract} 
Semantic communication aims to convey meaning for effective task execution, but differing latent representations in AI-native devices can cause semantic mismatches that hinder mutual understanding. This paper introduces a novel approach to mitigating latent space misalignment in multi-agent AI-native semantic communications. In a downlink scenario, we consider an access point (AP) communicating with multiple users to accomplish a specific AI-driven task. Our method implements a protocol that shares a semantic pre-equalizer at the AP and local semantic equalizers at user devices, fostering mutual understanding and task-oriented communication while considering power and complexity constraints. To achieve this, we employ a federated optimization for the decentralized training of the semantic equalizers at the AP and user sides. Numerical results validate the proposed approach in goal-oriented semantic communication, revealing key trade-offs among accuracy, communication overhead, complexity, and the semantic proximity of AI-native communication devices.
\end{abstract}
\smallskip
\begin{IEEEkeywords}
Semantic Communication, semantic equalization, latent space alignment, MIMO, Federated Learning. 
\end{IEEEkeywords}

\vspace{.15 cm}
\section{Introduction and Motivation}
\label{sec:intro}

Traditional communication systems emphasize accurate transmission of bits or symbols. However, the increasing number of connected devices and data-intensive applications is rapidly straining network capacity \cite{de2021survey}. Time-critical tasks like autonomous driving and smart surveillance demand reliable, low-latency AI services while managing energy, bandwidth, and computation efficiently. Despite advances, current wireless systems increasingly consume bandwidth and energy to meet data demands, underscoring the need for a new, more efficient communication paradigm. 
Semantic communication is emerging as a key enabler for 6G networks by representing raw symbols in a compressed, task-relevant form, thereby reducing bandwidth use and latency \cite{strinati2024goal,gunduz2022beyond}. Recent research builds on this foundation, exploring joint source-channel coding \cite{gunduz2022beyond}, semantic extraction  \cite{kountouris2021semantics}, goal-oriented system design \cite{di2023goal}, semantic reasoning \cite{thomas2023reasoning}, and generative AI \cite{barbarossa2023semantic}.

Artificial intelligence (AI) is a key enabler of semantic communications, where deep neural networks (DNNs) embed raw data into low-dimensional semantic features for transmission \cite{xie2021deep, gunduz2022beyond}. Effective communication requires a shared latent space between transmitter and receiver.
However, devices may encode the same information into different latent representations due to independent training procedures or architectural discrepancies. This misalignment introduces \textit{semantic noise}, which can significantly impair mutual understanding between communicating agents. Such discrepancies are not exceptions but rather common occurrences—particularly in multi-vendor environments where parties are unwilling or unable to share models, training datasets, or other proprietary assets. In these cases, joint end-to-end training of encoding networks or the exchange of DNN models is often infeasible due to privacy concerns and intellectual property constraints. As a result, an effective semantic alignment mechanism becomes essential to ensure consistency and interoperability across heterogeneous systems. This challenge, known as \textit{semantic channel equalization}, has prompted various alignment solutions, including relative representations (RRs)
\cite{moschella2022relative}, linear mappings \cite{lahner2024direct}, and optimal transport (OT)-based on global invariance \cite{alvarez2019towards, kuhnel2021latent}. Recent works have further addressed semantic noise and channel impairments via OT-codebooks \cite{sana2023semantic} and dynamic RR-based schemes \cite{fiorellino2024dynamic, huttebraucker2024relative}.
Despite growing interest in semantic alignment, most studies target single-link scenarios, leaving the multi-user case largely unexplored.

\noindent\textbf{Contributions.} The goal of this work is to address the latent space mismatch in a downlink semantic communication scenario, where an AP communicates with a set of AI-native users, each equipped with its own DNN model. The proposed method implements a protocol featuring a shared semantic pre-equalizer at the AP and local semantic equalizers at user devices. This architecture facilitates mutual understanding, implements semantic compression, and enables task-oriented communication under power and complexity constraints. Assuming linear semantic equalization modules at the transmitter and receiver, the optimization problem is cast as a block-convex program and is numerically solved using a federated alternating direction method of multipliers (ADMM) framework \cite{boyd2011distributed}, enabling decentralized training of both the semantic pre-equalizer and equalizers. Finally, a message exchange protocol is introduced to reduce communication overhead and latency, while preserving the privacy of users' personal latent spaces. As far as we know, this is the first aligner designed specifically for broadcast multi-user semantic communication scenarios available in the literature. Numerical results confirm the robustness of the proposed method in a multi-user goal-oriented semantic communication scenario for image classification.

\section{System Model}

\begin{figure}[t]
    \centering
    \includegraphics[width=\columnwidth, trim=10bp 8bp 8bp 5bp, clip]{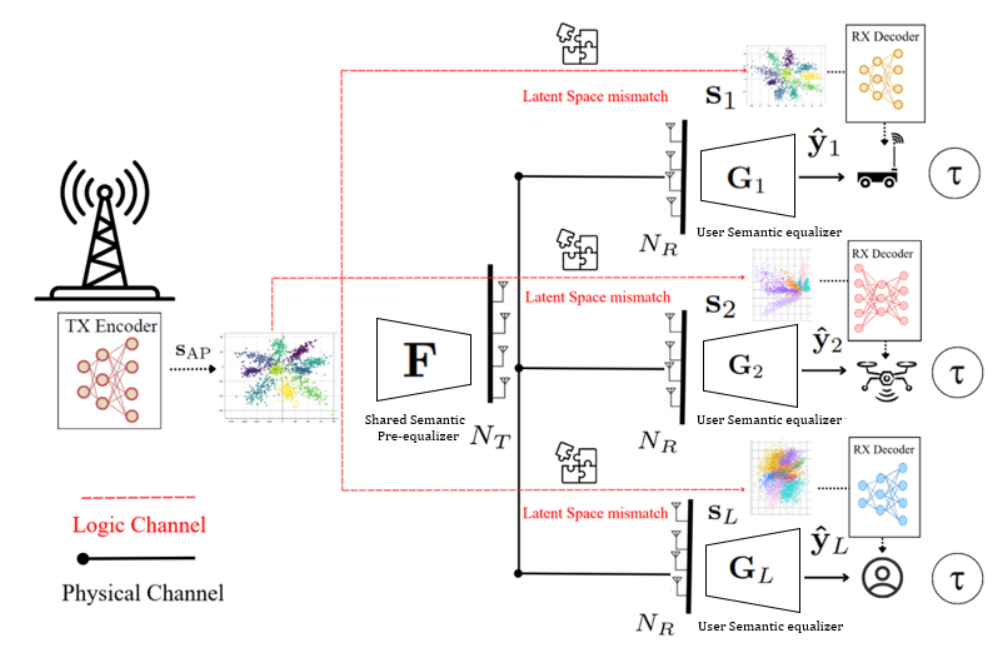}
    \vspace{-0.25 cm}
    \caption{\small Overview of the proposed system model. }
    \label{fig:system_mode}
\end{figure}

We consider a downlink system where an AP and 
$L$ users, all endowed with pre-trained DNNs, communicate semantically via transmitted and interpreted latent representations. A pictorial sketch is illustrated in Fig. \ref{fig:system_mode}.  Let $\mathbf{s}_{\rm AP}\in \mathbb{R}^d$ be the vector of semantic features extracted at the TX side from a data point $\mathbf{z}\in \mathbb{R}^q$. The set of all semantic vectors $\mathbf{s}_{\rm AP}$ represents the TX semantic latent space at the AP. Every user $l$ is trained to interpret a different encoding scheme than the one used by the AP, i.e., it requires the reception of a different semantic feature vector, say $\mathbf{s}_{l}\in \mathbb{R}^{m_l}$ (corresponding to $\mathbf{z}$), to correctly interpret the transmitted message or effectively perform a given task (e.g., classification). The set of all semantic vectors $\mathbf{s}_{l}$ represents the RX semantic latent space at each user $l\in\{1,\ldots,L\}$. Latent space mismatches introduce semantic noise, requiring equalization to ensure mutual understanding between the AP and users.
Our approach to semantic equalization exploits the presence of MIMO wireless channels between the AP and the users. Assuming, without loss of generality, that $d$ is even, we proceed by pairing the first half of the semantic features in $\mathbf{s}_T\in \mathbb{R}^d$ with the second half to form complex symbols, yielding an input vector $\x\in\mathbb{C}^\frac{d}{2}$. Then, assuming the AP is endowed with $N_T$ antennas, we exploit a \textit{semantic pre-equalizer} that maps the complex vector $\x\in\mathbb{C}^\frac{d}{2}$ into the vector $\overline{\mathbf{x}}\in\mathbb{C}^{K N_T}$ to be transmitted over $K$ wireless channel usages. The semantic pre-equalizer at the AP implements a learnable transformation represented by the function $f: \mathbb{C}^\frac{d}{2}\to \mathbb{C}^{K N_T}$. This transformation 
enables the semantic compression of the TX latent space (since typically $\frac{d}{2} \gg K$), with a compression factor given by  
$\displaystyle\zeta = \frac{K}{d/2}.$
For each user $l\in\{1,\ldots,L\}$, we consider the presence of a MIMO flat Rayleigh fading channel described by the matrix $\overline{\mathbf{H}}_l\in\mathbb{C}^{N_T\times N_R}$, where $N_R$ denotes the number of antennas at the RX side, assumed to be the same for all users.
For the sake of simplicity, we assume that all channels remain constant over the duration of $K$ consecutive transmissions. Additionally, in our simulations, we consider the case where perfect channel state information (CSI) is available at both TX and RX, while noting that TX may also operate without CSI.
Then, at every user $l\in\{1,\ldots,L\}$, we have a \textit{semantic equalizer} that maps the received symbols into a complex vector $\hat{\y}_l\in\mathbb{C}^\frac{m_l}{2}$, via the learnable transformation $g_l: \mathbb{C}^{K N_R}\to \mathbb{C}^{\frac{m_l}{2}}$. Overall, for every user $l$, the considered semantic MIMO communication channel can be compactly written as:
\begin{align}\label{eq:problem_setting}
    \hat{\y}_l = g_l(\H_l f(\x) + \n_l), \quad l=1,\ldots,N,
\end{align}
where, $\H_l =\I_K \otimes \overline{\H}_l\in\mathbb{C}^{KN_R\times KN_T}$, and $\n_l\in\mathbb{C}^{K N_R}$ denotes the noise vector at the $l$-th user, which follows a complex Gaussian distribution $\mathcal{CN}(\mathbf{0},\boldsymbol{\Sigma}_n)$ for all $l\in\{1,\ldots,N\}$. Finally, at every user $l$, the complex latent vector $\hat{\y}_l\in \mathbb{C}^{\frac{m_l}{2}}$ in (\ref{eq:problem_setting}) is converted into an $m_l$-dimensional real latent vector, say $\hat{\mathbf{s}}_l$, by inverting the halving operation done at the TX side.  Our aim is to act on the learnable transformations $f$ (at the BS side, shared among all users) and $\{g_l\}_{l=1}^L$ (at the user side) in (\ref{eq:problem_setting}) to perform the best possible semantic alignment between the AP and the users' latent spaces. This can be obtained by minimizing the (semantic) distance between the spaces composed by the vectors $\mathbf{s}_l$ and $\hat{\mathbf{s}}_l$, for all $l=1,\ldots,L$, over a limited set of available latent vectors, acting as \textit{semantic pilots} for logic-channel estimation.
In the sequel, we will illustrate the proposed federated strategy for latent space alignment. 
\section{Federated Latent Space Alignment}
We consider a semantic pre-equalizer function $f$ modeled as a linear transformation, represented by matrix $\F\in\mathbb{C}^{K N_T\times \frac{d}{2}}$; similarly, users' equalization functions $\{g_l\}_{l=1}^L$ are modeled by matrices $\G_l\in\mathbb{C}^{\frac{m_l}{2}\times K N_R}$, with $l\in\{1,\ldots,L\}$. Also, w.l.o.g., we consider the application of pre-whitening to standardize the covariance of the transmitted symbols ($\x$). Under these assumptions, the channel model (\ref{eq:problem_setting}) boils down to:
\begin{align}\label{eq:linear_observation}
  \hat{\y}_l
    = \G_l\H_l \F \x + \G_l\n_l,
    \quad l=1,\ldots,L .
\end{align}
Our optimization criterion aims at minimizing the (semantic) distance between the signals given by model \eqref{eq:linear_observation} and the target latent spaces of all users $l\in\{1,\ldots,L\}$, with respect to the (shared) linear semantic pre-equalizer $\F$ and the (local) equalizers $\{\G_l\}_{l=1}^L$. To this aim, let us assume that to have a training set of semantic pilots composed of $n$ examples $(\x_i,\y_{i,l})$, for $i\in\mathcal{T}_r$ and $l=1,\ldots,L$. Although various distance metrics can be applied within this framework, we opt for mean-squared error in what follows, as it offers a straightforward yet effective measure of latent space discrepancies. Mathematically, the problem is cast as an empirical risk minimization:
\begin{align}
    \min_{\mathbf{F}, \{\mathbf{G}_l\}_{l=1}^L} & \quad 
    \frac{1}{L n} \sum_{l=1}^L  \sum_{i \in \mathcal{T}_r}  \mathbb{E}\left\| \mathbf{y}_{i,l} - \mathbf{G}_l \left( \mathbf{H}_l \mathbf{F} \mathbf{x}_{i} + \mathbf{n}_l\right) \right\|_F^2 \notag \\
    \text{s.t.} & \quad \text{tr}(\mathbf{F} \mathbf{F}^H) \leq P_T
\label{eq:objective}
\end{align}
where the constraint in (\ref{eq:objective}) imposes a maximum budget $P_T$ on the power transmitted by the AP. Let us now define $\X\in\mathbb{C}^{\frac{d}{2}\times n}$ be the matrix that collects all training samples, where each column $\x_i$ represents an individual input,
and $\Y_l\in\mathbb{C}^{\frac{m}{2}\times n}$ be the matrix containing the corresponding latent column vectors $\y_{i,l}$ at the $l$-th user,
for all $l=1,\ldots,L$. Then, leveraging the zero-mean property of the noise components $\{\n_l\}_{l=1}^L$, the objective function of \eqref{eq:objective} can be recast as:
\begin{align}
    \min_{\mathbf{F}, \{\mathbf{G}_l\}_{l=1}^L} & \; \frac{1}{Ln} \sum_{l=1}^L  \left\| \mathbf{Y}_l - \mathbf{G}_l \mathbf{H}_l \mathbf{F} \mathbf{X}\right\|_F^2 
    +  \sum_{l=1}^L \text{tr} \left(\mathbf{G}_l \mathbf{\Sigma}_n \mathbf{G}_l^H\right)   \notag \\
    \text{s.t.} & \quad  \text{tr}(\mathbf{F} \mathbf{F}^H) \leq P_T
    \label{eq:matrix_form_original}
\end{align}
Problem (\ref{eq:matrix_form_original}) is non-convex, but enjoys a block-convex structure in the single variables, once having fixed the others. Thus, in the sequel, we will leverage a federated ADMM optimization framework to iteratively optimize the shared semantic pre-equalizer and the local users' equalizers. 

\subsection{Federated ADMM derivation}

As a first step, we recast  (\ref{eq:matrix_form_original})  in an equivalent manner, introducing the auxiliary variable $\mathbf{Z}$ and the set 
$\Omega=\{ \mathbf{Z} \,|\, {\rm tr}(\mathbf{ZZ}^H) \leq P_T \}.$
Then, we equivalently get:
\begin{align}
    \min_{\mathbf{F}, \{\mathbf{G}_l\}_{l=1}^L} & \; \frac{1}{Ln} \sum_{l=1}^L  \left\| \mathbf{Y}_l - \mathbf{G}_l \mathbf{H}_l \mathbf{F} \mathbf{X}\right\|_F^2 +I_{\Omega}(\mathbf{Z})
      \notag \\
    & +  \sum_{l=1}^L \text{tr} \left(\mathbf{G}_l \mathbf{\Sigma}_n \mathbf{G}_l^H\right)     \notag\\
    \text{s.t.} & \quad  \F-\mathbf{Z}=\mathbf{0}
    \label{eq:matrix_form_original2}
\end{align}
ADMM iteratively minimizes the (scaled) augmented Lagrangian of problem (\ref{eq:matrix_form_original2}) given by:
\begin{align}
L_{\rho}(\mathbf{F},\mathbf{G},\mathbf{Z},\mathbf{U}) &= \frac{1}{Ln} \sum_{l=1}^L  \biggl[ \left\| \mathbf{Y}_l - \mathbf{G}_l \mathbf{H}_l \mathbf{F} \mathbf{X}\right\|_F^2 +I_{\Omega}(\mathbf{Z}) \nonumber \\
&\hspace{-1cm} + \operatorname{tr} \bigl( \sum_{l=1}^L\mathbf{G}_l \mathbf{\Sigma}_n \mathbf{G}_l^H \Bigr) \biggr] + \rho \left\|\mathbf{F}-\mathbf{Z}+\mathbf{U} \right\|_F^2
\label{eq:lagrangian_}
\end{align}
with respect to primal variables, while minimizing it with respect to dual variables \cite{boyd2011distributed}; here, $\mathbf{U}$ is the (scaled) Lagrange multiplier enforcing the constraint in ($\ref{eq:matrix_form_original2}$), and $\rho$ is a positive parameter.
Now, we derive the single steps of ADMM.

\vspace{.1cm}
\noindent\textbf{The $\G_l$ step.} The Lagrangian (\ref{eq:lagrangian_}) is separable over the $\{\mathbf{G}_l\}_{l=1}^L$. Setting the gradient of  (\ref{eq:lagrangian_}) w.r.t. $\mathbf{G}_l^H$ to zero, and fixing $\F$ at time $t$, the update of $\G_l$ at time $t+1$ reads as: 
\begin{equation}
    \begin{aligned} 
        \mathbf{G}_l^{(t+1)} &= \mathbf{Y}_l (\mathbf{H}_l \mathbf{F}^{(t)} \mathbf{X})^H 
          \left( (\mathbf{H}_l \mathbf{F}^{(t)} \mathbf{X})(\mathbf{H}_l \mathbf{F}^{(t)} \mathbf{X})^H + n \mathbf{\Sigma}_n \right)^{-1}
        \label{eq:g_step_admm}
    \end{aligned}
\end{equation}

\noindent\textbf{The $\F$ step.}
With $\G$, $\Z$ and $\U$ fixed, the update for $\F$ at iteration $t+1$ can be obtained setting the gradient of (\ref{eq:lagrangian_}) w.r.t. $\mathbf{F}$ to zero. The solution depends on federated information that comes from all users $l=1,\ldots,L$, and can be expressed as:
\begin{equation}
  \mathbf{F}^{(t+1)}= \frac{1}{L} \sum_{l=1}^{L} \hat{\mathbf{F}}_l   
  \label{eq:global_F_update}
\end{equation}
where the single variables $\{\hat{\mathbf{F}}_l\}_{l=1}^L$ can be found as the solutions of the subproblems: 
\begin{align}
\hat{\F}_l =
&\, \arg\min_{\F_l} \;\frac{1}{n} ||\Y_l-\G_l^{(t+1)}\H_l\F_l\X||_F^2 \nonumber\\
&\;\;+\rho||\F_l-\Z^{(t)}+\U^{(t)}||^2_F \label{eq:f_local_admm}
\end{align}
for $l=1,\ldots, L$. Setting the gradient of (\ref{eq:f_local_admm}) to zero, we get:
\begin{align}
&(\mathbf{G}^{(t+1)}_l\mathbf{H}_l)^H (\mathbf{G}^{(t+1)}_l\mathbf{H}_l)\mathbf{F}_l (\mathbf{X}\mathbf{X}^H) + n \rho \F_l \nonumber\\
&\quad\;\; -n\rho(\Z^{(t)} - \U^{(t)})- (\G_l^{(t+1)}\H_l)^H \Y_l\X^H = \mathbf{0}.
\label{eq:F_l objective to zero}
\end{align}
Now, letting
\begin{align}
 & \A_l=(\G_l^{(t+1)}\H_l)^H(\G_l^{(t+1)}\H_l), \label{eq:A}\\
 & \B=\X\X^H,\label{eq:B}\\
 & \C_l=n\rho(\Z^{(t)}-\U^{(t)})+(\G_l^{(t+1)}\H_l)^H\Y_l\X^H, \label{eq:C}
\end{align}
we can compactly recast (\ref{eq:F_l objective to zero}) as:
\begin{equation}\label{eq:Sylvester}
     \mathbf{A}_l\mathbf{F}_l\mathbf{B} + n \rho \mathbf{F}_l =  \mathbf{C}_l, 
\end{equation}
which is a Sylvester equation that can be efficiently solved by Bartels-Stewart algorithm.
Otherwise, to find the closed form solution of (\ref{eq:Sylvester}), exploiting 
$\text{vec}(\A_l\F_l\B)=(\B^H\otimes \A)\text{vec}(\F_l)$
and solving for $\text{vec}(\F_l)$, after easy algebra we obtain:
\begin{equation}
\hat{\F}_l = \text{vec}^{-1}\left((\B^H\otimes \A_l + n\rho \mathbf{I} )^{-1}\text{vec}(\C_l)\right)  \label{eq:f_step_local_final}
\end{equation}
%

\vspace{.1cm}
\noindent\textbf{The $\Z$ step.} Minimizing (\ref{eq:lagrangian_}) w.r.t. to $\mathbf{Z}$ leads to:
\begin{align}\label{eq:Z_update}
\Z^{(t+1)} 
&= \arg\min_\Z \,||\F^{(t+1)}+\U^{(t)}-\Z||^2_F +I_{\Omega}(\mathbf{Z}) \nonumber\\
&= \text{Proj}_\Omega(\F^{(t+1)}+\U^{(t)}), 
\end{align}
where $\text{Proj}_\Omega (\cdot)$ denotes the projection operator onto the set $\Omega$. Easy algebra shows that the optimal solution of (\ref{eq:Z_update}) reads as:
\begin{equation}
   \Z^{(t+1)} = \frac{1}{1+\hat\lambda}(\F^{(t+1)}+\U^{(t)}),\label{eq:z_step_final}
\end{equation}
where $\hat\lambda =\max\left(0,\sqrt{\frac{\text{tr}(\hat{\Z}\hat{\Z}^H)}{P_T}}-1\right)$ with $\hat{\Z}=\F^{(t+1)}+ \U^{(t)}$.

\noindent\textbf{The $\U$ step.} 
The Lagrange multiplier update reads as:
\begin{align}
\U^{(t+1)} = \U^{(t)} + \F^{(t+1)} - \Z^{(t+1)}
\label{eq:u_step}
\end{align}


\subsection{Protocol for Federated Implementation}
\label{sec:protocol}
In this section, we summarize the protocol outlining the main steps of the proposed federated ADMM algorithm, along with the data exchange between the AP and the devices. This protocol is specifically designed to exchange the variables with a reduced payload, while maintaining private users' latent space representations. The main steps are given by:
%
\begin{itemize}
    \item[i)] An initial \textit{handshaking} step, in which users submit requests to the AP, which determines the number of devices involved in the alignment process. 
    \item[ii)] To update locally their  semantic decoder $\mathbf{G}_l^{(t+1)}$ by (\ref{eq:g_step_admm}), every user $l$ receives from the AP the variable $\mathbf{S}_l^{(t)} = \mathbf{H}_l\mathbf{F}^{(t)}\mathbf{X}$ if the AP has CSI and $\mathbf{S}_l^{(t)} = \mathbf{F}^{(t)}\mathbf{X}$ otherwise.
    

    \item[iii)] To update $\mathbf{F}^{(t+1)}$ at transmitter side using (\ref{eq:f_step_local_final}), the $l$-th user forwards $\mathbf{A}_l$ in (\ref{eq:A}) and 
    $\mathbf{P}_l=(\mathbf{G}_l^{(t+1)}\mathbf{H}_l)^H\mathbf{Y}_l$ to the AP.
   To enforce privacy constraint, 
   typical of a federated paradigm, 
   pre-whitening is applied to $\mathbf{Y}_l$. 
   This step ensures privatization of local latent representation $\mathbf{s}_l$.
     \item[iv)] Updates of $\mathbf{Z}^{(t+1)}$ and $\mathbf{U}^{(t+1)}$ are computed by the AP.
     \item[v)] Return to step (ii), and repeat until iteration $T$.
\end{itemize}
All the main steps of the proposed Federated ADMM for semantic alignment are summarized in Algorithm $1$.

\begin{algorithm}[t]
\caption{\textbf{Algorithm 1 :} Federated ADMM for Semantic Alignment}
\label{alg:admm_algo}
\begin{algorithmic}[1]
\STATE \textbf{Input:} \( \F^{(0)} \sim \mathcal{CN}(0,1) \), \( \{\Z^{(0)} , \U^{(0)}\} = \mathbf{0} \), $\rho>0$.
\STATE \textbf{Output:} Final values \( \F^{(T)}, \G^{(T)}, \Z^{(T)}, \U^{(T)} \).\\
\STATE Perform \textit{handshaking};
\FOR{each iteration \( t = \{1, \dots, T \}\)}
    \STATE The AP transmits $\mathbf{S}_l^{(t)}$ to all users $l=1,\ldots,L$;\\
    The users update locally  $\G^{(t+1)}_l$ as in \eqref{eq:g_step_admm};
    \STATE Users send $\mathbf{A}_l$ and $\mathbf{P}_l$ to the AP;\\
    The AP updates $\F^{(t+1)}$ as in \eqref{eq:global_F_update};
    \STATE The AP updates $\Z^{(t)}$ as in \eqref{eq:z_step_final};
    \STATE The AP updates $\U^{(t)}$ as in \eqref{eq:u_step}.
\ENDFOR 
\label{algo:protocol}
\end{algorithmic}
\end{algorithm}
\section{Numerical Results}
This section provides numerical results to assess the performance of the proposed algorithm in a multi-user semantic communication system, transmitting latent representations to perform an image classification task. We consider the CIFAR-10 dataset, comprising 60000 $32\times32$ color images distributed across 10 classes (6000 images per class). Among them, 42500 images were used for training, 7500 for validation, and 10000 for testing, with classification across 10 labels as the downstream task $\tau$. 
Latent representations are produced by the backbone of pre-trained models chosen from the \textit{timm} Python library and are summarized in Table~\ref{tab:models}.
We consider square MIMO Rayleigh fading channels with unitary variance. 
All displayed results are averaged across seeds:\{27,42,100,123,144,200\}, considering  $\rho=1$, transmitter power constraint $P_T$ equal to $1$ and iterations $T=30$.\footnote{\href{https://github.com/SPAICOM/multi-agent-semantic-alignment.git}{https://github.com/SPAICOM/multi-agent-semantic-alignment.git}}

\noindent \textbf{Baselines.} We compare our methodology with baselines that directly transmit vectors $\x$ of the TX latent space while performing semantic alignment and MIMO channel equalization in a disjoint fashion. Specifically, leveraging SVD channel decomposition as $\overline{\mathbf{H}}_l = \U_l\S_l\V_l^H$, we set $\mathbf{G}_l = \mathbf{1}_K^T \otimes \left( \mathbf{\Sigma}_l^H\mathbf{\Sigma}_l + \frac{1}{\text{SNR}_l} \mathbf{I}_{N_R} \right)^{-1}\left(\mathbf{U}_l\mathbf{\Sigma}_l\right)^H$ for all $l=1,\ldots,L$, where SNR$_l$  represents the Signal-to-Noise Ratio in dB (SNR) at the $l$-th user; the semantic pre-coder is obtained by solving:
\begin{align}
\mathbf{F} = \arg\min_{\mathbf{F}} \quad & \frac{1}{Ln} \sum_{l=1}^L \mathbb{E} \Big\|\mathbf{X} - \mathbf{G}_l\Big(\mathbf{H}_l \mathbf{F}\mathbf{X} + \mathbf{n}_l\Big)\Big\|_F^2 \nonumber \\ 
\text{s.t.} \quad & \operatorname{tr}(\mathbf{F}\mathbf{F}^H) \le P_T.
\label{eq:F_objective_baseline}
\end{align}
We consider two different transmitting strategies: (i) "First-$K$" baseline, which transmits the first $2 \cdot K \cdot N_T$ features of $\x \in \mathbb{R}^d$, and (ii) "Top-$K$"  baseline, that transmits the largest $K$ features based on their $\textit{l}_1$ norm. In the latter, along with their features, corresponding indices are also sent, used by the receivers to reconstruct the original signal, assuming perfect index reconstruction. After signal transmission, semantic alignment is performed by users by solving a least square problem:
\begin{align}\label{pr:least_square}
\min_{\mathbf{Q}\in\mathbb{C}^{m_l\times d}} ||\mathbf{Q}_l \X-\Y_l||_F^2.
\end{align}
for $l=1,\ldots,L$, where $\mathbf{Q}_l$ represents the alignment matrix. Furthermore, we benchmarked our approach against a multiplexing scheme in which each $l$-communication link is endowed with an independently optimized semantic pre-coder, under the assumption of negligible multi-user interference.

\begin{table}[t!]
    \centering
    \caption{\small Models Configuration}
    \label{tab:models}
    \begin{tabular}{ll}
        \textbf{Group} & \textbf{Models} \\
        \hline
        AP & vit\_tiny\_patch16 \\ 
        \hline
        Users & efficientvit\_m5, levit\_128, rexnet\_100, \\
                      & vit\_small\_patch\{16,32\}, vit\_base\_patch\{16,32\}, \\
              & mobilenet\_v3\_small\_\{075,100\}, mobilenet\_v3\_large\_100\\ 

        \hline
        Heterogeneous & vit\_\{small,base\}\_patch16\_224, \\
                        & mobilenetv3\_small\_075, efficientvit\_m5 \\ 
        \hline
        Homogeneous & mobilenetv3\_small\_\{075,100\},\\
                    & rexnet\_100,  mobilenetv3\_large\_100\\ 
        \hline
    \end{tabular}
\end{table}

\begin{figure}[t]
    \centering    \includegraphics[width=\columnwidth, trim=10bp 8bp 5bp 5bp, clip]{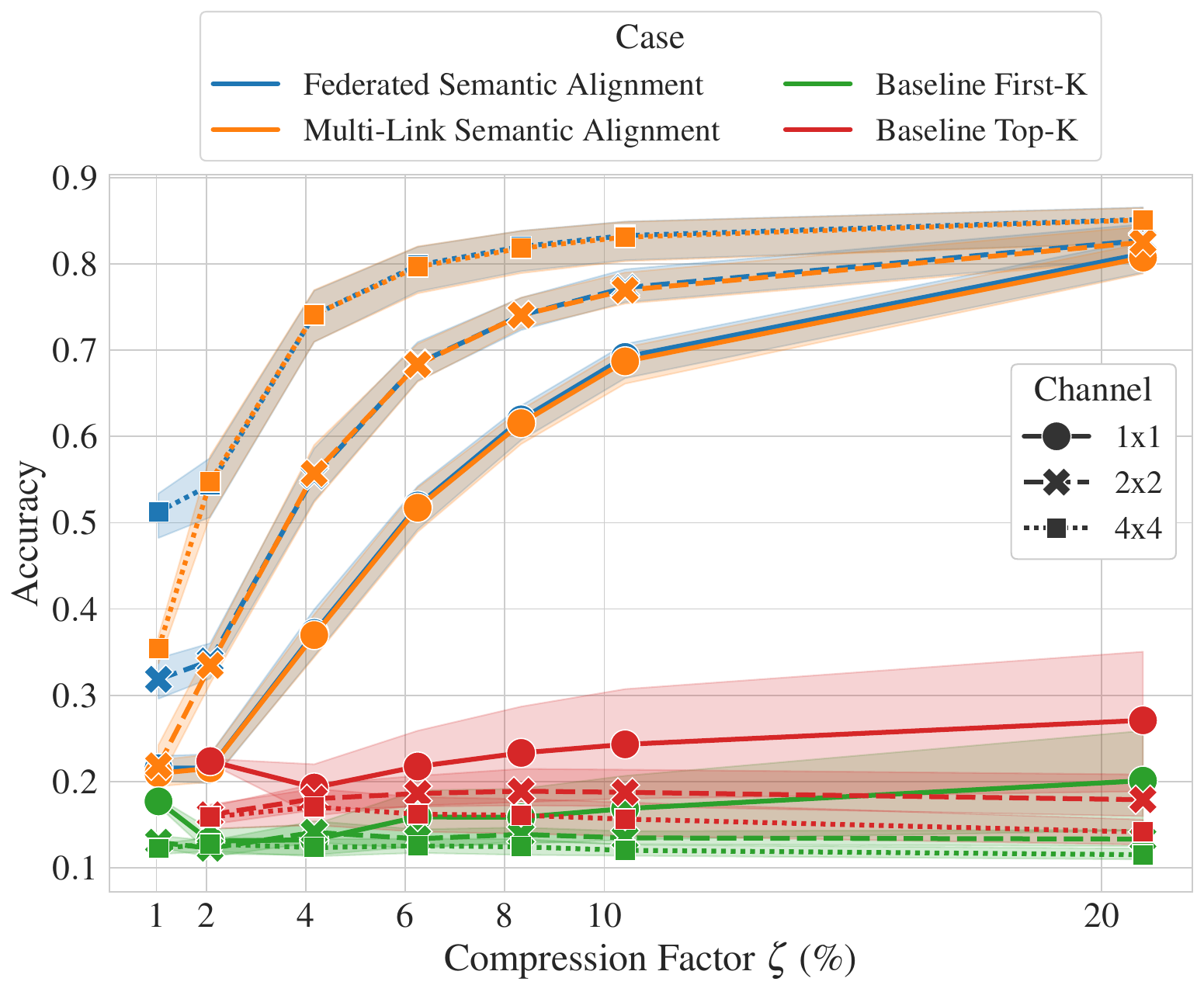}
    \vspace{-0.1cm}
    \caption{\small Accuracy vs. $\zeta$ with SNR=20 dB.}
    \label{fig::compression_fact}
\end{figure}
\begin{figure}[t]
    \centering    \includegraphics[width=\columnwidth, trim=7bp 9bp 7bp 7bp, clip]{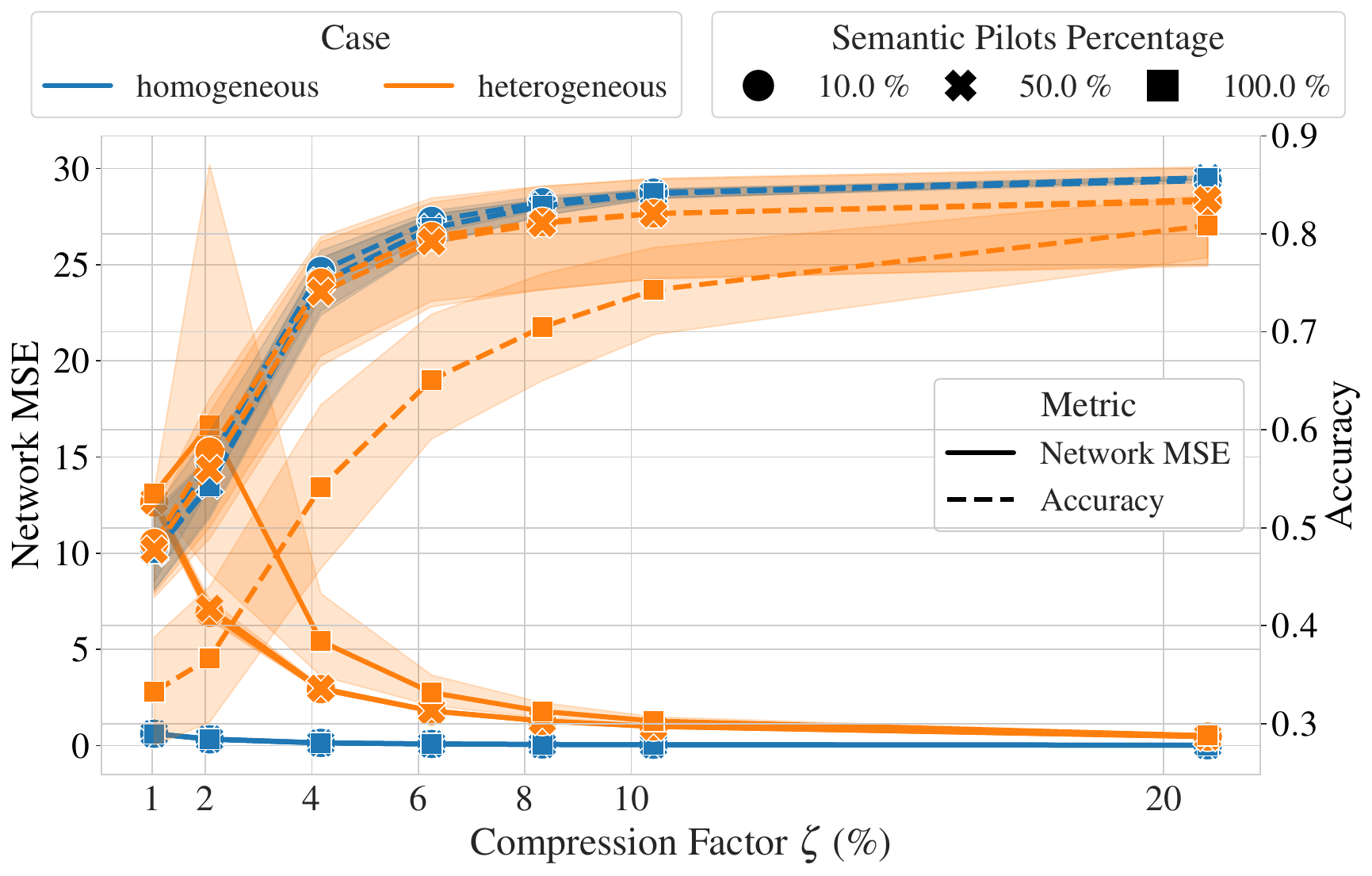}
    \vspace{-0.1cm}
\caption{Alignment struggle as Network MSE and Accuracy vs. $\zeta$ for a 4$\times$4 MIMO channel with SNR=20 dB.}
    \label{fig::alignment struggle}
\end{figure}

\noindent \textbf{Results.} As a first example, Figure \ref{fig::compression_fact} presents the average downstream task accuracy versus compression factor $\zeta$ for various MIMO antenna configurations, wherein our Federated approach is benchmarked against the aforementioned baselines. The simulation considers a scenario with 10 users and SNR equal to 20 dB. As we can notice from Fig. \ref{fig::compression_fact}, the proposed method significantly outperforms the First-$K$ and Top-$K$ baseline strategies. Interestingly, the proposed approach exhibits enhanced performance also with respect to the multi-link baseline configuration, which employs different semantic pre-equalizers over $L$ independent parallel channels. 
In fact, mitigating semantic discrepancies in the latent spaces generated by heterogeneous models and the AP, the shared semantic pre-equalizer ensures robust performance even at high compression rates, underscoring the strong semantic alignment capabilities of our approach, without the need for having separate semantic equalizers and channels for each user.
Semantic alignment performance is affected by the degree of semantic similarity among the latent spaces produced by different models. Intuitively, the AP is better able to align users whose latent representations are semantically similar. To quantitatively evaluate this effect, Fig. \ref{fig::alignment struggle} reports the network mean-squared error (MSE) between the latent spaces of the AP and users, and the task accuracy, as a function of the compression factor $\zeta$, using our proposed method under two distinct settings, displayed in Table \ref{tab:models}: (i) a homogeneous setup, where four users employ different but semantically similar encoders; and (ii) a heterogeneous setup, where the four users are split into two groups with significantly different latent space structures. We also consider three different percentages of semantic pilots available for training, i.e., $10\%$, $50\%$, and $100\%$. As shown in Fig. \ref{fig::alignment struggle}, in both setups the network mean squared error (MSE) decreases as $\zeta$ increases, whereas, conversely, the accuracy improves. As expected, the homogeneous configuration results in significantly lower network MSE, along with more stable and higher task accuracy compared to the heterogeneous setting. This is due to more favorable semantic alignment conditions in the homogeneous case. Moreover, reducing the percentage of semantic pilots can facilitate model alignment and enhance overall task accuracy. However, this benefit holds only down to a certain threshold, below which alignment deteriorates and performance drops. This reveals an interesting trade-off between the number and selection of semantic pilots and overall system performance. We plan to explore this aspect in future work.

\section{Conclusions}
This paper proposes a federated latent space alignment framework for multi-user semantic communications. By combining a shared semantic pre-equalizer at the AP with personalized semantic equalizers on user devices, the framework effectively addresses the challenge of aligning diverse neural representations in the 
presence of semantic noise. The proposed methodology hinges on a federated ADMM method enabling decentralized training of both the semantic pre-equalizer and the user-side equalizers. Extensive numerical evaluations with heterogeneous user models validate the approach, demonstrating its ability to balance key performance indicators such as accuracy, communication overhead, and computational complexity. Future research includes optimal semantic pilot selection, data-driven semantic clustering, and managing user coexistence in interference-limited scenarios. Key challenges involve interference-aware alignment and dynamic user grouping based on latent space compatibility. Addressing these aspects will be essential for realizing efficient multi-user semantic communication.


\bibliographystyle{IEEEtran}
\bibliography{references}
\end{document}